\documentclass[12pt]{article}%
\usepackage{fancyhdr}%
\usepackage{amsmath}%
\usepackage{amsfonts}%
\usepackage{amssymb}%
\usepackage{graphicx}

\begin{document}

\title{Superlinear amplitude amplification}
\author{Lov K. Grover \thanks{Research was partly supported by NSA\ \&\ ARO under
contract DAAG55-98-C-0040.}\\\textit{lov.grover@gmail.com}\\
}
\date{}
\maketitle

\begin{abstract}
Quantum search/amplitude amplification algorithms are designed to be able to
amplify the amplitude in the target state linearly with the number of
operations. Since the probability is the square of the amplitude, this results
in the success probability rising quadratically with the number of operations.
This paper presents a new kind of quantum search algorithm in which the
amplitude of the target state, itself increases quadratically with the number of
operations. However, the domain of applications of this is much more limited
than standard amplitude amplification.

\end{abstract}

\section{Background}

Quantum searching was invented to speed up the searching process in databases.
It was realized by Hoyer et al\cite{amplitude-amp-hoyer} and independently
by me \cite{ampt-amp-grover} that this searching was a special case of
amplitude amplification whereby the amplitude in a target state could be
amplified linearly with the number of operations. This realization
considerably increased the power of the algorithm, no longer was it limited to
database searching but was applicable to a host of physics and computer
science problems. In fact it gave a square-root speedup for almost any
classical probabilistic algorithm.

The idea behind this speedup was realized later on to be a two dimensional
rotation through which the state-vector got driven from the source to a target
state through a sequence of small rotations in the two dimensional space
defined by the source and the target state.

This is easily seen by considering the basic transformation:\ $-UI_{s}%
U^{-1}I_{t}U$ say $V{\Large .}$ Then if we calculate $V_{ts}$, by definition
of the $I_{t}\;\&\;I_{s}$ operations, it easily follows that
\begin{subequations}
\label{0}%
\begin{equation}
V_{ts}=(-UI_{s}U^{-1}I_{t}U)_{ts}=3U_{ts}-4\left|  U_{ts}\right|  ^{2}%
\approx3U_{ts}\label{amptamp}%
\end{equation}

Note that this is true for any unitary $U.$ It stays true if we replace $U$ by
$V$ which yields:%
\end{subequations}
\[
(-VI_{s}V^{-1}I_{t}V)_{ts}=3V_{ts}%
\]
Substituting for $V$ as $-UI_{s}U^{-1}I_{t}U$ in  (\ref{recursion}) and
$V_{ts}$ from (\ref{amptamp}), it follows that:%
\[
(-UI_{s}U^{-1}I_{t}UI_{s}U^{-1}I_{t}UI_{s}U^{-1}I_{t}UI_{s}U^{-1}I_{t}%
U)_{ts}\approx9U_{ts}%
\]

Similarly by recursing multiple times, we can prove the transformation:
$U(-I_{s}U^{-1}I_{t}U)_{ts}^{p}\approx(2p+1)U_{ts}$ to be true for large $p.$

\section{Quadratic Amplitude Amplification}

This paper gives a new kind of amplitude amplification in which the amplitude
in the target grows ;quadratically with the number of iterations. Instead of
$\ $choosing the basic transformation to be $V{\Large =}-UI_{s}U^{-1}I_{t}U,$
we choose $V$ to be $-U^{-1}I_{t}I_{s}U.$ It follows by using the definitions
of $I_{s}$ \&\ $I_{t},$ that
\begin{equation}
V_{ts}=\left(  -U^{-1}I_{t}I_{s}U\right)  _{ts}=2U_{ss}U_{ts}^{\ast}%
+2U_{tt}^{\ast}U_{st}\label{recursion}%
\end{equation}
In case $U_{ss}$ $\approx$ $U_{tt}^{\ast}$ and $U_{st}\approx U_{ts}^{\ast},$
then $V_{ts}\approx4U_{ss}U_{ts}.$Unlike the recursion equation of the
previous transformation which only depended on $U_{ts}$, this equation depends
on both $U_{tt}$ and $U_{ss}$ and even $U_{st}.$ So we need to investigate how
$U_{tt}$ and $U_{ss}$ vary in successive recursions.

Consider $V_{ss}$. Again assuming $U_{st}\approx U_{ts}^{\ast}$ and
$U_{ss}\approx U_{tt}^{\ast}$%
\[
V_{ss}=\left(  -U^{-1}I_{t}I_{s}U\right)  _{ss}=-1+2\left|  U_{ss}\right|
^{2}-2\left|  U_{ts}\right|  ^{2}%
\]
Note that if we denote $U_{ss}=\left(  1-\delta\right)  $, and $V_{ss}$ by
$\left(  1-\gamma\right)  $ assuming all terms to be real and neglecting
\ $2\left|  U_{ts}\right|  ^{2}$ on the RHS, the above equation may be written as:%

\[
\gamma\approx4\delta
\]

Therefore $V_{ss}$ stays close to 1 for approximately$\frac{\ln\frac{1}%
{\delta}}{\ln4}$ recursions. In i recursions, provided $U_{tt}\approx1$,
$U_{ts}$ rises by a factor of approximately $4^{i};$ therefore in
$\frac{\ln\frac{1}{\delta}}{\ln4}$ recursions $U_{ts}$ rises by approximately
a factor of $\frac{1}{\delta}.$ The number of queries is approximately
$2^{\left(  \frac{\ln\frac{1}{\delta}}{\ln4}\right)  }$ which is
$\frac{1}{\sqrt{\delta}},$ as expected the amplification of $U_{ts}$ is
quadratic in the region when $U_{ss}$ is approximately 1.

\section{Example - U\ is the Inversion about Average Operation}

Consider the situation when s, the starting state is an arbitrary basis state
and U is the inversion about average transformation. Then, assuming there to
be N states to be searched, $U_{ss}$ is $-1+\frac{2}{N}$ and $U_{ts} $ is
$\frac{2}{N}.$ Then analyzing the sequence of operations for a few steps-

\begin{itemize}
\item $U=WI_{0}W$

\item $-U^{-1}I_{t}I_{s}U=-\underbrace{WI_{0}W}\;I_{t}I_{s}\underbrace
{WI_{0}W}=W\;I_{0}\;W\;\left(  I_{t}I_{s}\right)  \;W\;I_{0}\;W$

\item
\begin{align}\label{seq}
U^{-1}I_{t}I_{s}U\;I_{s}I_{t}\;U^{-1}I_{t}I_{s}U  & =\underbrace{WI_{0}%
W}\;I_{t}I_{s}\underbrace{WI_{0}W}\;I_{s}I_{t}\;\underbrace{WI_{0}W}%
\;I_{t}I_{s}\underbrace{WI_{0}W}\nonumber\\
& =\cdots W\;I_{0}\;W\;\left(  I_{t}I_{s}\right)  \;W\;I_{0}\;W\;\left(
I_{t}I_{s}\right)  \;W\;I_{0}\;W%
\end{align}
\end{itemize}

Looks something like the search algorithm, which is:
\[
=\cdots W\;I_{t}\;W\;I_{t}\;W\;I_{s}\;W\;I_{t}\;W\;I_{s}\;W
\]

However, any similarity is superficial, as we discuss in the following
section, this algorithm is \emph{not} a rotation of the state vector in
two-dimensional Hilbert space.

Nevertheless, the dynamics of the algorithm are fairly simple to understand
and analyze iteratively: The state just before an inversion about average is
described by 3 parameters, the amplitudes in the target state, source state
and that in the other states.%

\begin{figure}
[ptb]
\begin{center}
\includegraphics[
natheight=7.499600in,
natwidth=9.999800in,
height=2.8202in,
width=3.7507in
]%
{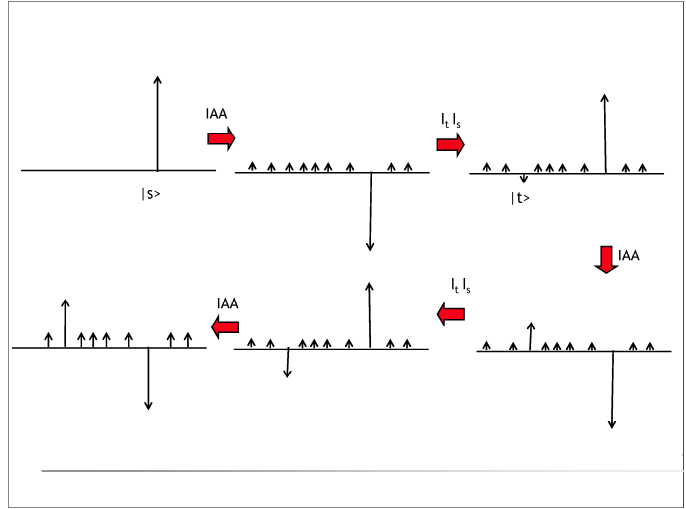}%
\caption{This describes two iterations of the new quantum search algorithm. As
described above, the U\ transform is the Inversion about Average operation
(IAA). }%
\end{center}
\end{figure}

The evolution is obtained by the following equations (A(t) denotes the average
amplitude over all states):%

\begin{align*}
\Delta A(t)  & =2\frac{S(t)}{N}-2\frac{T(t)}{N}\\
\Delta S(t)  & =-2A(t)\\
\Delta T(t)  & =2A(t)
\end{align*}

Going to the continuous limit and solving this system of differential
equations gives the amplitude in the target state as $\frac{1}{2}-\frac{1}%
{2}\cos\left(  \frac{2\sqrt{2}t}{\sqrt{N}}\right)  .$ Therefore in
$t=\frac{\pi\sqrt{N}}{2\sqrt{2}}$ iterations, the amplitude in the target
state becomes unity.

\subsection{Observations}

\begin{description}
\item The number of iterations required for searching with certainty is
$\sqrt{2}$ times more than required by the search algorithm.

\item The variation of the amplitude in the target state is $\frac{1}%
{2}-\frac{1}{2}\cos\left(  \frac{2\sqrt{2}t}{\sqrt{N}}\right)  .$ In the
initial stages (when $t$ is close to 0), the amplitude varies as
$\frac{2t^{2}}{N}$. As expected, the rate of increase is quadratic. However,
once the probability in the target state become significant (also affecting
$U_{ss})$, the quadratic nature of the increase is destroyed.

\item The algorithm of this paper may be useful in applications where the
basic $U_{ts}$ that needs to be amplified is small (in the above example where
the U\ transform was the inversion about average, $U_{ts}$ was only
$\frac{2}{N}$ - whereas in the search algorithm it is about $\frac{1}{\sqrt
{N}}.$

\item It is possible that there would exist applications where a few
applications of this algorithm provided the driving transform for amplitude
amplification algorithms. That way, we would get the quadratic speedup plus
the flexibility of the amplitude amplification algorithms.
\end{description}

\subsection{This is not the search algorithm}

One might be tempted to conclude that the above algorithm was a variant of the
search algorithm because, overall, it gave a square-root speedup; also it
consists of similar sequences of unitary transformations (\ref{seq}). However,
that is not the case.

The chief characteristic of the search algorithm and all its variants
(amplitude amplification algorithms) was a rotation of the state vector in
appropriately defined two dimensional space. The algorithm of this paper needs
more than two dimensions to operate in. To see this consider the basic
recursion equation (\ref{recursion}) used to develop the algorithm:
$V_{ts}=\left(  -U^{-1}I_{t}I_{s}U\right)  _{ts}=2U_{ss}U_{ts}^{\ast}%
+2U_{tt}^{\ast}U_{st}.$ Given large $U_{ss}$ \&\ $U_{tt}$ and $U_{st}\approx
U_{ts}^{\ast},$ we had argued that $V_{ts}$ was amplified significantly in
each recursion. In order to satisfy this condition needs more than two
dimensions. This is because if there were only two dimensions it would follow
from unitarity of $U$ that $2U_{ss}U_{ts}^{\ast}+2U_{tt}^{\ast}U_{st}$=0 (any
two columns of a unitary matrix are orthogonal) - therefore, additional
dimensions are necessary.

\section{ Conclusion}

The above algorithm gives a quadratic amplification under certain
conditions.The quadratic amplification offers something new beyond the search
algorithm, even though it is not as universally applicable. To borrow a term
from analog amplifiers: this only has a limited dynamic range - outside of
this range it has to be supplemented by other more robust algorithms.

Just as \cite{quantum-search} (\cite{fixed-point-quantum}%
,\cite{fixed-point-tulsi}), this algorithm provides yet another tool in the
quantum algorithm designer's toolkit. Whereas, ( \cite{quantum-search}) is
independently useful to design quantum algorithms, the fixed point algorithms
\&\ the algorithm of this paper may be  useful in combination with \ other
algorithms - ( (\cite{fixed-point-quantum},\cite{fixed-point-tulsi})) to
improve the robustness and the algorithm of this paper  to increase the
amplification in selected  ranges.

As described in the \textit{Observations} section, the algorithm of this paper
may be useful in conjunction with the standard quantum search algorithm. This
is somewhat similar in spirit to applications where the search algorithm is
combined with a classical algorithm. One such example is the counting
algorithm of \cite{amplitude-amp-hoyer} where one gets round the cyclical
nature of the search algorithm by making appropriately timed observations
(which is the classical algorithm). The counting algorithm is not usually
looked at this way, but in the context of robustness versus speed, it is
insightful to look upon it as a combination of classical and quantum search algorithms,%

\begin{figure}
[ptb]
\begin{center}
\includegraphics[
trim=0.000000in 1.456422in 0.000000in 0.000000in,
natheight=7.499600in,
natwidth=9.999800in,
height=2.2779in,
width=3.7507in
]%
{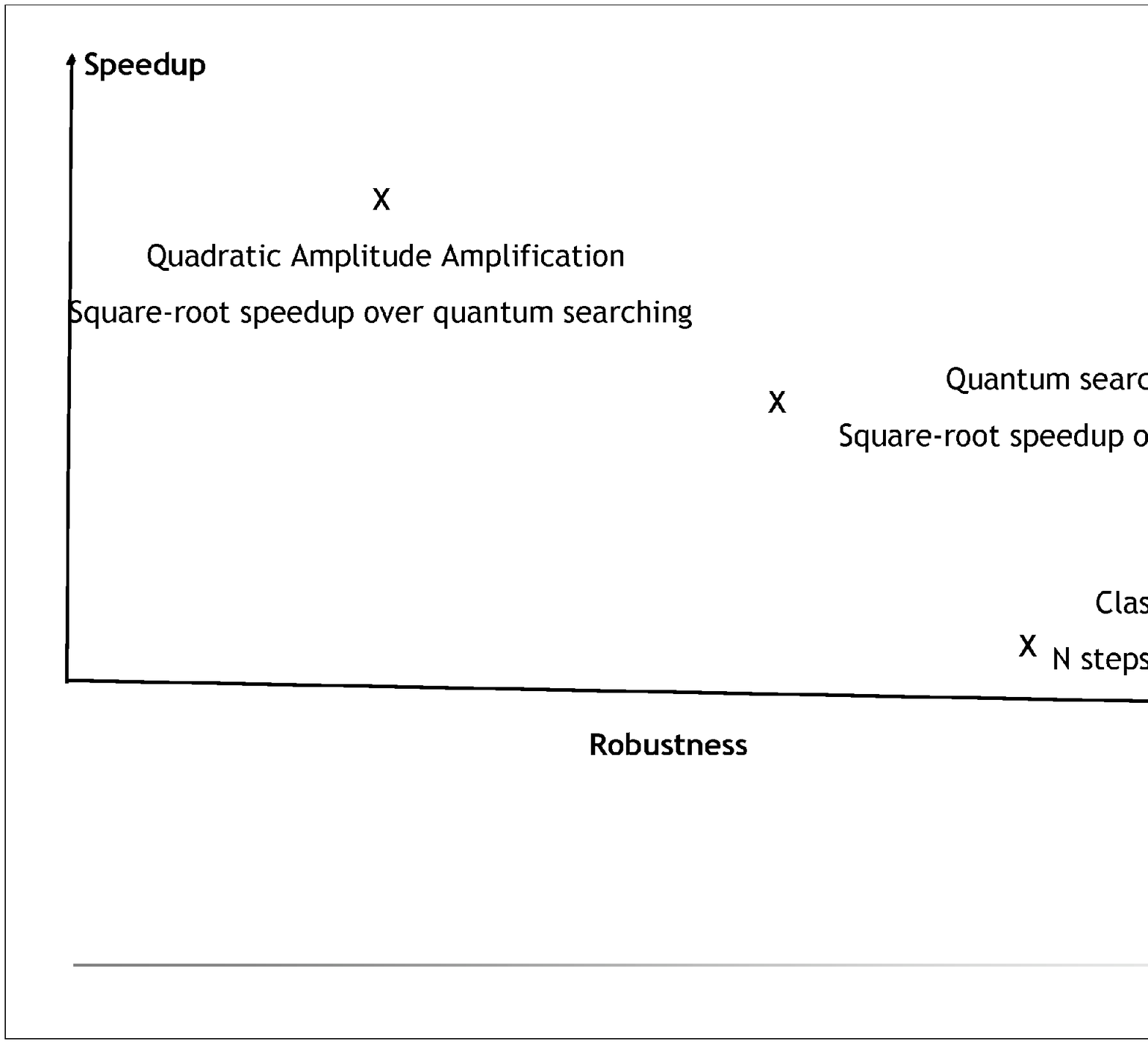}%
\caption{Hierarchy of search algorithms - The quantum search algorithm
\ attained a square-root speedup over classical but the price paid was more
sensitivity. The present algorithm provides a square-root speedup over quantum
searching, but it still more sensitive.}%
\end{center}
\end{figure}

\end{document}